\documentclass[letterpaper,12pt]{article}
\renewcommand{\baselinestretch}{1.5}
\usepackage{times}
\usepackage[margin=1in]{geometry}

\usepackage{graphicx}
\DeclareGraphicsExtensions{.pdf,.jpeg,.png}
\usepackage{cite}
\usepackage{booktabs}
\usepackage[symbol*]{footmisc}
\usepackage{slashbox}

\usepackage{footnote}

\usepackage{amsmath}
\usepackage{amsfonts}
\usepackage{titlesec}
\usepackage{algorithmic}
\usepackage{algorithm}
\titlelabel{\thetitle.\quad}
\titleformat*{\section}{\normalfont\bfseries}
\titleformat*{\subsection}{\normalfont\bfseries}
\titleformat*{\subsubsection}{\normalfont\bfseries}
\titleformat*{\paragraph}{\normalfont\bfseries}
\titleformat*{\subparagraph}{\normalfont\bfseries}

\newcommand*{\myfnsymbolsingle}[1]{%
  \ensuremath{%
    \ifcase#1
    \or 
      *%
    \or 
      **
    \or 
      ***
    \or 
      ****
    \or 
      *****
    \else 
      \@ctrerr
    \fi
  }%
}

\begin{document}
\date{}

\title{FUZZY LOGIC CONTROL FOR MIXED CONVENTIONAL/BRAKING ACTUATION MOBILE ROBOTS }

\author{Walelign~Nikshi\footnote{Dr. Nikshi is a Control Systems Engineer at Icon build, Austin, TX, 78745 USA e-mail: wallee394@gmail.com.},
        ~Randy~C.~Hoover\footnote{Dr. Hoover is with the Department of Electrical and Computer Engineering, South Dakota School of Mines and Technology, Rapid City, SD, 57701 USA e-mail: Randy.Hoover@sdsmt.edu.},
        ~Mark~D.~Bedillion \footnote{Dr. Bedillion is with the Department of Mechanical Engineering, Carnegie Mellon University, Pittsburgh, PA, 15213 USA e-mail: mbedillion@cmu.edu.},
        ~Saeed~Shahmiri
        \footnote {Mr. Shahmiri is Lab Coordinator and M.Sc. graduated with the Department of Electrical Engineering, South Dakota School of Mines and Technology, Rapid City, SD, 57701, USA e-mail: Saeed.Shahmiri@sdsmt.edu}
        \\and~Jeremy~Simmons \footnote{Mr. Simmons is a Ph.D student at the Department of Mechanical Engineering, University of Minnesota, Minneapolis, MN, 55455 USA e-mail: simmo536@umn.edu.}
}



\maketitle

\thispagestyle{empty}

\noindent
{\bf\normalsize Abstract}\newline

The use of conventional actuators in robotic systems (electric motors in particular), while often offering advantages in terms of flexibility and controllability, suffer from primary actuator failure, due to unexpected complexities in their environment, which can lead to loss of controllability. Conventional actuators can impose disadvantages on mechanical complexity, weight, and cost. Here, the Mixed conventional/braking Actuation Mobile Robot (MAMR), a new mobile robot platform, is proposed to tackle such drawbacks in actuation and explore the use and control of braking actuation. This platform substitutes the drive motors used in Ackermann steering with brakes that have only two states, ON and OFF. 
Additionally, the conventional drive wheels are replaced by a single, omni-directional wheel that only supports a driving force in the robot’s longitudinal direction. The ability of braking actuators in providing controllability under actuator failure is one of the primary motivations of this work. To validate the reliability and accuracy of MAMR approach, this paper studies the design of such robotic systems, the design and synthesis of fuzzy logic controllers along with the experimental assessments of these controllers in real-time. The experimental tests point out the controller performance enhancement using fuzzy logic controllers and MAMR. \vspace{2ex}
   
\noindent
{\bf\normalsize Key Words}\newline
Mobile robots, conventional actuation, braking actuation, mixed conventional/braking actuation, parking control, fuzzy logic control.

\section{Introduction}

{M}{obile} robots have received much interest from the engineering community in the last few decades due to their many applications.
Examples include, factory automation (e.g., automated guided vehicles), military operations (e.g., unmanned ground reconnaissance vehicles), health care (e.g., pharmaceutical delivery), search and rescue, security, and in the household (e.g., floor cleaning and lawn mowing)~\cite{schneier2015literature}.
While the research spans many large functional areas, locomotion is the most common and primary function of all mobile robots to accomplish a particular task in their environments. 
These environments, structured and/or unstructured, pose significant challenges because of their inherent uncertainty and complexity.
As a result, mobile robots often require an application-specific locomotion strategy. 
Mobile robots are typically characterized by two classes of locomotion~\cite{siegwart2011introduction}: locomotion systems that mimic biological counter-parts, and locomotion systems that use actively powered wheels.
Locomotion systems that mimic biological counter-parts include swimming~\cite{dudek2005visually, dudek2007aqua}, flying~\cite{bouabdallah2007design,cutler2012design}, walking~\cite{machado2006overview, deshmukh2006robot}, skating~\cite{iverach2012ice}, and others~\cite{stepan2009acroboter, wang2008biological, menon2004gecko,art:Jun2018,art:Li2017,art:Chen2017, andani2018fuzzy}.
Locomotion systems that use actively powered wheels are characteristic of wheeled mobile robots (WMR)~\cite{morin2008motion,de2001control}. 
In the space of all mobile robots, WMRs are widely used in applications with low mechanical complexity and energy consumption whereas biologically inspired robots, such as legged robots, are more apt in nonstandard, or unstructured, environments such as stairs, heaps of rubble, generally uneven and harsh terrain~\cite{tzafestas2013introduction}. 
\begin{figure}[!t]
\centering
\includegraphics[scale=1.3]{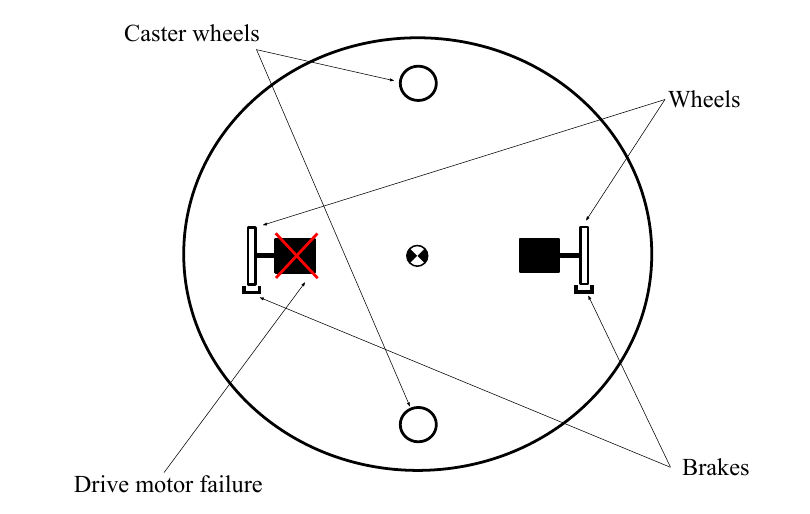}
\caption{Differential drive robot with brakes to regain controllability under drive motor failure.}
\label{regain_controllability}
\end{figure}
\raggedbottom

All mobile robots, regardless of the locomotion strategy they use, require some sort of actuation for influencing each degree of freedom. 
There are two general classes of actuators used in robotic applications: conventional actuators and braking actuators.
Examples of conventional actuators include electric motors, hydraulic and pneumatic actuators~\cite{mavroidis2005}. 
While robots can employ any of these actuators, electric motors are the most commonly used actuators because of their well-established control methods, low volumetric requirements, high precision, and a readily available power source. 
A primary drawback of electric motors is the low torque output-weight ratio that elicits a need for gearing to reduce speed and increase torque. 
This can result in an excessive backlash, friction, wear, and undesired compliance, and manifest in inaccuracy, poor dynamic response, and poor torque control capability.
To tackle these drawbacks in robotic applications that require human interactions such as exoskeleton and prosthetics, Series Elastic Actuators (SEA) are used \cite{pratt1995series,pratt2004series}.
This will help to control the force at the joints precisely and increase the shock tolerance and compliance with its environment. 

Another type of actuation system used in robotics is braking actuation. 
Most robots that use brakes use continuously variable brakes such as those found in automobiles~\cite{litman2014autonomous}.
However, this work considers brakes with only ON/OFF states because this can enable simple, cheap, and small braking design as an add-on to robots that do not include brakes.
The use of two-state braking actuation was studied in~\cite{bedillion2014distributed} where a grid of brakes in an inclined plane influenced the position and orientation of an object.
Framstad \textit{et al.}~\cite{framstad2015control} studied the control of a mobile robot with three ON/OFF brakes as its only form of actuation as it maneuvered on an inclined plane. 

One of the primary motivations for studying braking actuators in mobile robotics is that they could allow the system to regain controllability under actuator failure. 
For example, consider a differential drive mobile robot with brakes added along the axis of the conventional wheels as shown in Fig.~\ref{regain_controllability}. 
If one of the actuators fails during operation, the objective can still be achieved in contrast to the traditional differential drive mobile robot.
If the brake attached to the failed drive motor is activated, the robot will spin about the braked wheel point~\cite{nikshi2018trajectory}.  
With the brake inactive, it will drive in a spiral.
As a result, the target can be achieved by judiciously switching between the two braking states.
\begin{figure}[!t]
\centering
\includegraphics[scale=1.4]{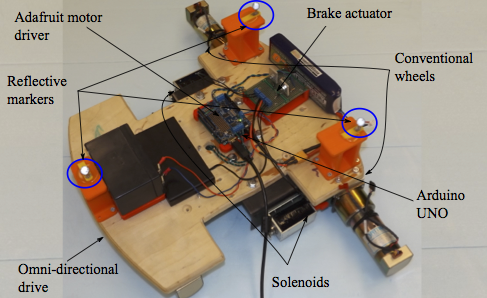}
\caption{The prototype of the MAMR.}
\label{MAMR_II_Jour}
\end{figure}
\raggedbottom

In addition to their use in the case of actuator failure, braking actuators have some advantages that might make them the primary actuation strategy in some cases.
Other advantages include: 
1) reduced weight in the case of a high degree of freedom mobile robots, 
2) ease of actuation,
3) reducing the overall cost as brakes cost less than conventional actuators, and
4) helping in miniaturization of mobile robotic systems as braking actuators have smaller space requirements than conventional actuators.
For example, consider the solenoid used in Fig. \ref{MAMR_II_Jour} as opposed to a DC motor with the same actuation force as outlined in Table~\ref{compare_solenoid_DCmotor}.
As can be seen, while providing the same actuation force, solenoids are smaller, lighter, and less costly than DC motors making the robotic system lighter, easier to actuate, more compact, and less costly. 
One disadvantage however, is that brakes introduce discontinuities into the dynamics making controller design more complex. 
\begin{table}[!t]
\centering
\caption{\selectfont Comparison of Solenoids and DC motors}
\label{compare_solenoid_DCmotor}
\begin{tabular}{| c|c|c| }
\hline 
& \multicolumn{1}{c|}{\begin{tabular}[x]{@{}l@{}} Solenoid\\ (ZHO-1364s-36A13)  \end{tabular} } & \multicolumn{1}{c|}{\begin{tabular}[x]{@{}l@{}} DC motor\\ (PD3665-12-5-BF)  \end{tabular} }  \\ \hline 
\begin{tabular}[x]{@{}l@{}} Actuation \\force  \end{tabular} & $17.65N$         & \begin{tabular}[x]{@{}l@{}} $0.54Nm$  ($\sim 17N$ \\on a wheel radius of 35$mm$)  \end{tabular}  \\ \hline
 Size ($mm$)     & $64.4\times 38\times 30$   & $70\times 36\times 36$  \\ \hline
 Weight ($kg$)   & $0.301$          & $0.394$     \\ \hline
 Cost($\$$)      & $14.95$          & $148.4$      \\ \hline
\end{tabular}
\end{table}
\raggedbottom
%

This work combines the use of conventional and braking actuators to form a new mobile robot platform, the Mixed conventional/braking Actuation Mobile Robot (MAMR). 
A prototype of the MAMR considered in this paper is shown in Fig.~\ref{MAMR_II_Jour}.
The robot has three wheels: one omni-directional drive wheel in the back and two freely spinning conventional wheels with brakes in the front side of the robot. 
The details on the design of this robot are outlined in Section~\ref{experimental_setup}.
This configuration of the MAMR differs from the robot shown in Fig.~\ref{regain_controllability}, in that the driving and braking actuation occurs at different points.
The driving force comes from the omni directional drive wheel. 
The brakes are used for steering and stopping the robot.
For example, the robot steers to the left if the left brake is activated and to the right if the right brake is activated for the positive driving force.
The opposite is true if the driving force is negative.
The rate of steering depends on the magnitude of the driving force.  
The braking action works based on friction between the wheel and ground.
Friction is notoriously complex, and makes modeling more difficult, which in turn makes control design harder.  

The use of Fuzzy Logic Control (FLC) has been applied in many different application domains, a subset of which include tracking~\cite{nikshi2018trajectory}, navigation and control of robotic systems~\cite{art:zadah73,castillo2006fuzzy,hoover2005hybrid,hoover2005fusion,rashid2010fuzzy,andani2018fuzzy}.
Fuzzy logic control is advantageous due to its ability to incorporate expert knowledge into the control law via if-then heuristic rules.
In our initial work~\cite{simmons2016mechatronic, nikshi2016parking}, we presented simulation results illustrating a solution to the parking problem for the MAMR using FLC.  Fuzzy control was chosen as the control methodology for its ability to deal with model uncertainties and inaccuracies through expert knowledge,  e.g., those that arise from uncertain friction models~\cite{art:zadah73}. Robles et al. could manage the uncertainaty in controller performance using type-2 fuzzy logic approach In the current work we extend our previous work~\cite{simmons2016mechatronic, nikshi2016parking} in two distinct ways:  1)  the design and control of a new braking actuation system that places nonholonomic constraints on the MAMR dynamics (useful for physical realization and control of the system) and 2) through validation of the FLC experimentally via physical implementation as opposed to pure simulation.  
 
The remainder of the paper is organized as follows: 
Section~\ref{modeling} presents the mathematical modeling of the MAMR. 
Section~\ref{control} discusses the controller design for the parking problem. 
The prototype and experimental setup used for validation of the controller is presented in Section~\ref{experimental_setup}, followed by a discussion and comparison of simulation and experimental results in Section~\ref{experimental_result}. 
Finally, conclusions and directions for future work are presented in Section~\ref{conclusion_recommendation}.
\raggedbottom

\section{Mathematical Modeling}  \label{modeling}

\subsection{Assumptions} \label{assumptions}
\begin{figure}[!t]
\centering
\includegraphics[scale=1.5]{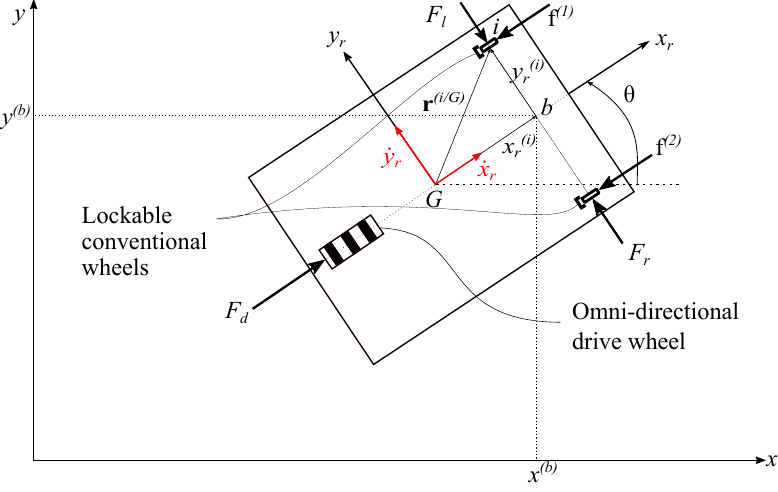}
\caption{The free body diagram of the MAMR.}
\label{CoordinateSystems_MAMR2_Jour}
\end{figure}

The dynamics of the MAMR are derived under the assumption that it is a rigid body operating on the special Euclidean group $SE(2) = \mathbb{R}^2 \times \mathbb{S}^1$.  As shown in Fig.~\ref{CoordinateSystems_MAMR2_Jour}, the global coordinate frame ($x, y$) is fixed in space and the local coordinate frame ($x_r, y_r$) is fixed at the center of mass of the robot.
Consequently, the position of the robot's center of mass in the global frame is given as $x$ and $y$, and in the local frame as $x_r$ and $y_r$.
The orientation of the robot remains the same in both frames and is given $\theta$.
Additional assumptions include:
\begin{enumerate}
\item The omni-directional wheel is a Swedish 90-degree wheel that exerts force $F_d$ along the center axis of the robot ($x_r$) only.
\item The braking actuators operate in two states, locked/unlocked (i.e. exerting friction or not exerting friction).
\item The omni-directional drive wheel rolls without slip.
\item The contact force at each braking actuator assumes a Coulomb friction model.
\item The force of friction of each braking actuator acts at a single point.
\item The brakes are in sliding contact with the ground, causing the robot to translate while rotating.
\item The braking force is sufficient to lock the wheel relative to the robot.
\item The center of mass is located at the geometric center of the three contact points and thus the weight is divided equally among them.
\end{enumerate}

The equations of motion for the MAMR are derived in the local coordinate frame using the Newton-Euler method and can be transformed into the global frame using a standard orthogonal transformation as
\begin{equation}
\textbf{R}(\theta) = \begin{bmatrix}
\cos\theta&-\sin\theta&0\\
\sin\theta&\cos\theta&0\\
0&0&1\\
\end{bmatrix}. 
\label{fig:RotationMatrix} 
\end{equation}
The relationship between the local and global coordinate frames is given as
\begin{equation}
\textbf{x} = {\textbf{R}(\theta)}{\textbf{x}_r},
\label{fig:CoordinateTransformation} 
\end{equation} 
where $\textbf{x}_r$ and $\textbf{x}$ are the local and global configuration vectors defined as $\textbf{x}_r=[x_r,\:y_r,\: \theta]^T$ and $\textbf{x}=[x,\:y,\:\theta]^T$, respectively. 
As shown in~\cite{watanabe1998control}, the same rotation matrix can be used to transform the states between the local and global coordinate frames with an appropriate definition of velocity and acceleration vectors. 
For the MAMR, the velocity and acceleration vectors are defined as 
\begin{equation}
 	\dot{\textbf{x}}_r=[\dot{x}_r,\: \dot{y}_r,\:\dot{\theta}]^T
    \label{eqn:MAMR_vel}
\end{equation}
and 
\begin{equation}
	\ddot{\textbf{x}}_r=[\ddot{x}_r - \dot{y}_r \dot{\theta},\: \ddot{y}_r + \dot{x}_r \dot{\theta},\: \ddot{\theta}]^T
    \label{eqn:MAMR_acc}
\end{equation}
in the local frame, respectively, where $\dot{x}_r$ is the velocity along the $x_r$-axis, $\dot{y}_r$ is the velocity along the $y_r$-axis, $\dot{\theta}$ is the angular velocity, $\ddot{x}_r$ is the acceleration along the $x_r$-axis, $\ddot{y}_r$ is the acceleration along the $y_r$-axis, and $\ddot{\theta}$ is the angular acceleration of the robot.
Similarly, the velocity and acceleration in the global coordinate frame are defined as  $\dot{\textbf{x}}=[\dot{x},\:\dot{y},\: \dot{\theta}]^T$ and $\ddot{\textbf{x}}=[\ddot{x},\: \ddot{y},\: \ddot{\theta}]^T$, respectively, where $\dot{x}$ is the velocity along the $x$-axis, $\dot{y}$ is the velocity along the $y$-axis, $\ddot{x}$ is the acceleration along the $x$-axis, and $\ddot{y}$ is the acceleration along the $y$-axis of the robot.

\subsection{Kinematic Constraints} \label{kinematics}
The motion of the MAMR is constrained in the lateral direction by the conventional wheels. This constraint amounts to forcing the velocity of point $b$ in Fig.~\ref{CoordinateSystems_MAMR2_Jour} in the $y_r$ direction, $\dot{y}^{(b)}_r$, to be zero, i.e., 
\begin{equation}
 \dot{y}^{(b)}_r = -\dot{x}^{(b)} \sin(\theta) + \dot{y}^{(b)} \cos(\theta) = 0,
  \label{NoslipConstraint_2}
\end{equation} 
where $\dot{x}^{(b)}$ and $\dot{y}^{(b)}$ are the velocity components of point $b$ in the global coordinate frame. 

From Fig.~\ref{CoordinateSystems_MAMR2_Jour}, one can derive the kinematic relationships between point $b$ and the center of mass, $G$, in the global coordinate frame as
\begin{equation}
  \begin{aligned}
  & x^{(b)} = x +  x^{(i)}_r  \cos(\theta),\\
  & y^{(b)} = y +  x^{(i)}_r \sin(\theta),\\
  \end{aligned}
  \label{position}
\end{equation} 
where $x^{(b)}$ and $y^{(b)}$ are the $x$ and $y$ positions of  point $b$ in the global coordinates, respectively, and $x^{(i)}_r$ is the position of brake $i$ with respect to the $x_r$-axis.
Differentiating~(\ref{position}) and solving for the velocity of point $G$ with respect to the velocity of point $b$ yields
\begin{equation}
  \begin{aligned}
  & \dot{x} = \dot{x}^{(b)} + x^{(i)}_r \dot{\theta} \sin(\theta),\\
  & \dot{y} = \dot{y}^{(b)} -  x^{(i)}_r \dot{\theta} \cos(\theta).\\
  \end{aligned}
  \label{Speed_c_b}
\end{equation} 
The velocity of $G$ in the global coordinate frame can be written in terms of the local velocity as
\begin{equation}
  \begin{aligned}
  & \dot{x} = \dot{x}_r \cos(\theta) - \dot{y}_r \sin(\theta),\\
  & \dot{y} = \dot{x}_r \sin(\theta) + \dot{y}_r \cos(\theta).\\
  \end{aligned}
  \label{Speed_c_Vu_Vw}
\end{equation} 
Equating~(\ref{Speed_c_b}) and (\ref{Speed_c_Vu_Vw}), and solving for the velocity of point $b$ in the global coordinate frame gives
\begin{equation}
  \begin{aligned}
  & \dot{x}^{(b)}= \dot{x}_r \cos(\theta) - \sin(\theta) ( \dot{y}_r + x^{(i)}_r \dot{\theta}), \\
  & \dot{y}^{(b)} =\dot{x}_r \sin(\theta) + \cos(\theta)(\dot{y}_r + x^{(i)}_r \dot{\theta}).\\
  \end{aligned}
  \label{Speed_b_Vu_Vw}
\end{equation} 
The global velocity components of point $b$ with respect to the velocity in the local coordinates can also be written as
\begin{equation}
  \begin{aligned}
  & \dot{x}^{(b)} = \dot{x}^{(b)}_r \cos(\theta) - \dot{y}^{(b)}_r  \sin(\theta),  \\
  & \dot{y}^{(b)} = \dot{x}^{(b)}_r  \sin(\theta) + \dot{y}^{(b)}_r  \cos(\theta). \\
  \end{aligned}
  \label{Speed_b_dot_xr}
\end{equation} 
where $\dot{x}^{(b)}_r$ and $\dot{y}^{(b)}_r$ are the $x_r$ and $y_r$ velocity components of point $b$, respectively.
Equating~(\ref{Speed_b_Vu_Vw}) and (\ref{Speed_b_dot_xr}) yields the expression for the lateral velocity of the point $b$ in the local coordinate frame as 
\begin{equation}
   \dot{y}^{(b)}_r  = \dot{y}_r + x^{(i)}_r\dot{\theta}. 
\label{lateral_motion}
\end{equation}
Using~(\ref{NoslipConstraint_2}) for the no lateral slip constraint, results in
\begin{equation}
  \dot{y}_r + x^{(i)}_r\dot{\theta}   = 0.
   \label{Lateral_contsraint}
\end{equation} 

%

\subsection{Kinetics} \label{kinetics}
The free body diagram with all the forces acting on the MAMR is shown in Fig.~\ref{CoordinateSystems_MAMR2_Jour}. 
The equations of motion are obtained using Newton-Euler in the local coordinate frame.
%
A discrete state $F^{(i)}\in \{0 \,\, ,\,\,1\}$ for brake $i,\ i=1,2,$ is introduced to handle the two states of the braking actuators where $F^{(i)} = 0$ for an inactive brake and $F^{(i)} = 1$ for an active brake. 
Following the assumed Coulomb friction model, the force of friction at brake $i$ is given by the vector $\textbf{f}^{(i)}$ such that
\begin{equation}
\textbf{f}^{(i)} = -{mg \mu^{(i)}_k}  F^{(i)} \frac{\textbf{v}^{(i)}}{3||\textbf{v}^{(i)}||},
 \label{FrictionForce}
\end{equation}
where $\mu^{(i)}_k$ is the coefficient of kinetic friction at braking point $i$ and $g$ is the gravitational constant.
The direction of the friction is established by $\textbf{v}^{(i)}$ as the velocity vector of the brake point $i$ being scaled by the inverse of its magnitude $||\textbf{v}^{(i)}||$.


The relationship between the linear and angular velocities for planar motion is given by
\begin{equation}
\textbf{v}^{(i)} = \textbf{v} + {\dot{\theta}}{\hat{k}} \times \textbf{r}^{(i/G)},
\label{eq_5}
\end{equation}
where $\textbf{v} = [\dot{x}_r,\: \dot{y}_r]^T$ is the velocity vector of $G$ in the local coordinate frame.
$\textbf{r}^{(i/G)} = [x^{(i)}_r,\: y^{(i)}_r,\: \theta]^T$ is the position vector of brake $i$ with $x^{(i)}_r$ and $y^{(i)}_r$ being the $x_r$ and $y_r$ components, respectively, and $\hat{k}$ is a unit vector perpendicular to the $x-y$ plane. 
Using the nonholonomic constraint relation given by~(\ref{Lateral_contsraint}),~(\ref{eq_5}) simplifies to 
\begin{equation}
\textbf{v}^{(i)} = \begin{bmatrix}
             \ \dot{x}_r - y^{(i)}_r \dot{\theta}  \\
             \  0 \\
             \end{bmatrix},
\label{BrakeVelocity_2}     
\end{equation}  
and its magnitude is given by 
\begin{equation}
||\textbf{v}^{(i)}|| = \sqrt{ (\dot{x}_r - y^{(i)}_r\dot{\theta} ) ^2 }.
 \label{velocity_Vector_II}
\end{equation} 
Therefore, the friction force vector given by~(\ref{FrictionForce}) reduces to 
  \begin{equation}
\textbf{f}^{(i)} = - \frac{{mg \mu^{(i)}_k} F^{(i)}}{3||\textbf{v}^{(i)}||}  \left( \begin{bmatrix}
       \ \dot{x}_r - \dot{\theta} y^{(i)}_r\\
       \  0\\
        \end{bmatrix}\right).
 \label{fig:FricForce2_mamrII}
\end{equation}  

The driving force, $F_d$, does not produce a moment about $G$ because it is acting along the $x_r$-axis, passing through $G$.
However, the friction forces given by~(\ref{FrictionForce}), that act at an offset of $y^{(i)}_r$ from the $x_r$-axis, produce moments about $G$.
The moment due to friction force, $M^{(i)}_\textbf{f}$, is given by
\begin{equation}
M^{(i)}_\textbf{f} = \frac{-mg\mu^{(i)}_k  F^{(i)} }{3||\textbf{v}^{(i)}||} \left( \textbf{r}^{(i/G)}\times \textbf{v}^{(i)} \right).
 \label{Moment_friction}
\end{equation} 
Substituting the local components of $\textbf{r}^{(i/G)}$ and $\textbf{v}^{(i)}$ vectors into~(\ref{Moment_friction}) results in
\begin{equation}
M^{(i)}_\textbf{f} = \frac{-mg \mu^{(i)}_k  F^{(i)} }{3||\textbf{v}^{(i)}||} \left(  \begin{bmatrix}
       \  x^{(i)}_r\\
       \  y^{(i)}_r\\       
        \end{bmatrix} \times  \begin{bmatrix}
       \  \dot{x}_r - y^{(i)}_r\dot{\theta} \\
       \  0\\       
        \end{bmatrix}  \right),
 \label{fig:momendueFric2_II}
\end{equation}
and computing the cross product, the moment due to friction force becomes
\begin{equation}
M^{(i)}_\textbf{f} = \frac{-mg \mu^{(i)}_k  F^{(i)}}{3||\textbf{v}^{(i)}||} \left( - y^{(i)}_r\dot{x}_r  + {y^{(i)}_r}^2 \dot{\theta}  \right).
 \label{fig:momendueFric3_II}
\end{equation}

\begin{figure}[!t]
\centering
\includegraphics[scale=1.4]{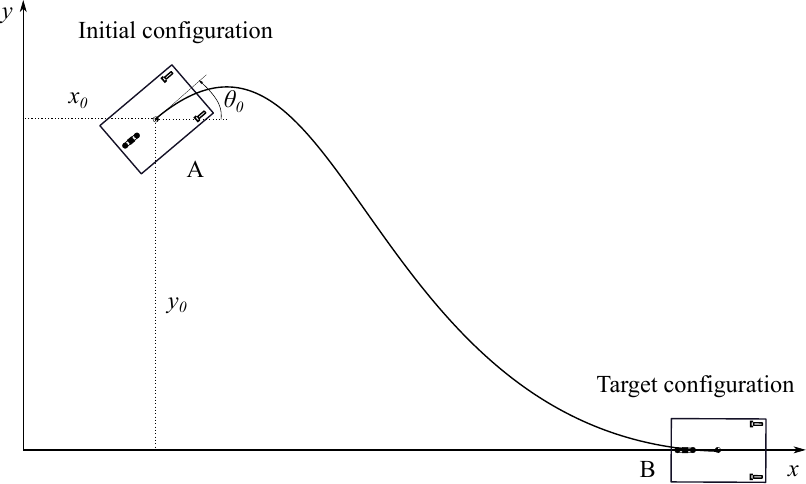}
\caption{Parking control problem formulation for the MAMR.}
\label{ParkingProblem_MAMR2}
\end{figure}
The reaction forces acting on the left and right wheels, $F_l$ and $F_r$, respectively, are acting along the axis of the wheels and are shown in Fig. \ref{CoordinateSystems_MAMR2_Jour}.
The magnitude of these forces is assumed to be unlimited, i.e. we assume that the nonholonomic constraint cannot be violated. 
The resultant force from these reaction forces, $R_f = F_r - F_l $, is also along this axis and will produce a moment about $G$.

Because the friction and driving forces are the only forces acting along the $x_r$-axis, the equation of motion along the $x_r$ direction is obtained as
\begin{equation}
    \ddot{x}_r = \frac{F_d}{m} -\sum_{i=1}^{2} \frac{g\mu^{(i)}_k F^{(i)} }{3||\textbf{v}^{(i)}||} \left(\dot{x}_r - y^{(i)}_r\dot{\theta}\right) + \dot{y}_r \dot{\theta}.
  \label{Dyanics_MAMRI}
\end{equation} 
On the other hand, the reaction forces are acting along the $y_r$ direction resulting in
\begin{equation}
m (\ddot{y}_r + \dot{x}_r\dot{\theta})= F_r - F_l = R_f.
 \label{fig:YDyanics_II_1}
\end{equation}
Solving for $\dot{y}_r$ and differentiating~(\ref{Lateral_contsraint}) with respect to time and substituting in~(\ref{fig:YDyanics_II_1}) yields
\begin{equation}
R_f = F_r - F_l = -m x^{(i)}_r \ddot{\theta} + m\dot{x}_r\dot{\theta}.
 \label{fig:reactionforce}
\end{equation}
  
The resultant reaction force, $R_f$, produces a counterclockwise moment about $G$ as
\begin{equation}
M_{R_f} =  x^{(i)}_r R_f,
 \label{fig:momendueReacForce}
\end{equation}
where $M_{R_f}$ is the moment due to reaction force $R_f$ about point $G$.
Substituting~(\ref{fig:reactionforce}) into (\ref{fig:momendueReacForce}), we get
\begin{equation}
M_{R_f} =  -m{x^{(i)}_r}^2 \ddot{\theta} + mx^{(i)}_r \dot{x}_r \dot{\theta}.
 \label{fig:momendueReacForce_2}
\end{equation}
Hence, the rotational dynamics becomes
\begin{equation}
\ddot{\theta} = \frac{1}{I + m{\alpha}^2} \sum_{i=1}^{2} \frac{mg \mu^{(i)}_k  F^{(i)} }{3||\textbf{v}^{(i)}||} \left(y^{(i)}_r \dot{x}_r  - {y^{(i)}_r}^2 \dot{\theta} \right) + \frac{m{\alpha}\dot{x}_r\dot{\theta}}{I + m{\alpha}^2}.
 \label{fig:RotatDyanics_II}
\end{equation}
Note that $\alpha$ replaced $x^{(1)}_r$ for terms that are outside the summation sign in (\ref{fig:RotatDyanics_II}).
Therefore, the equations of motion for the MAMR in the local coordinate frame are summarized as
\begin{align}
\ddot{x}_r =& \frac{F_d}{m} - \sum_{i=1}^{2} \frac{g\mu^{(i)}_k  F^{(i)} }{3||\textbf{v}^{(i)}||} \left(\dot{x}_r - y^{(i)}_r \dot{\theta}\right) + \dot{y}_r \dot{\theta},\\
\ddot{y}_r =& - \alpha\ddot{\theta},\\
\begin{split}
\ddot{\theta} =& \frac{1}{I + m{\alpha}^2} \sum_{i=1}^{2} \frac{mg \mu^{(i)}_k  F^{(i)} }{3||\textbf{v}^{(i)}||} \left( y^{(i)}_r \dot{x}_r  - {y^{(i)}_r}^2 \dot{\theta} \right) \\ &+ \frac{m \alpha\dot{x}_r\dot{\theta}}{I + m{\alpha}^2}.
\end{split}
\label{fig:2BareActiveDynamics_II}
\end{align}


\section{Control Design}
\label{control}

The control objective is to park the MAMR from a given initial configuration $\mathcal{C}_0 = [x_0,y_0,\theta_0] \in SE(2)$ to some desired final configuration $\mathcal{C}_d = [x_d,y_d,\theta_d] \in SE(2)$ in finite time, where the units on $\mathbb{R}^{2}$ are in meters and the units on $\mathbb{S}^1$ are in degrees.  In the current work, the final desired orientation is assumed to be zero degrees ($\theta_d = 0^\circ$) and the robot is to be parked along the $x$-axis (i.e. $y_d = 0$) as illustrated in Fig.~\ref{ParkingProblem_MAMR2}. 

The Fuzzy Logic Controller (FLC) has been explored to solve the parking control problem for the MAMR due to its effectiveness in controlling mechanical systems through the use of expert knowledge and its ability to deal with limited or uncertain information.
The details of design for its application to the MAMR parking control problem were first published in~\cite{nikshi2016parking} and are summarized here. 

\begin{figure}[!t]
\centering
\includegraphics[scale=1.5]{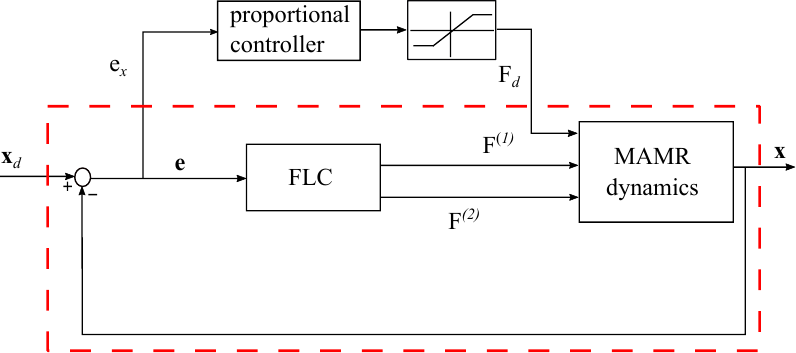}
\caption{The closed loop control architecture for parking control problem.}
\label{Feedback_parking_Jour}
\end{figure}

The controller follows a sequential architecture, depicted in Fig.~\ref{Feedback_parking_Jour}, to solve the parking problem in full.
The FLC first drives the MAMR to the desired $y$ position, $y_d$, and the desired orientation, $\theta_d = 0^\circ$. 
In an effort to simplify the design process, the driving force $F_d$ was kept constant thereby reducing the fuzzy rule base. 
A block diagram illustrating the FLC control is marked by the dotted line in Fig.~\ref{Feedback_parking_Jour}. 
Once $y_d$ and $\theta_d$ have been achieved via the FLC, the MAMR converges to the final desired position along the $x$-axis, $x_d$, via a proportional controller with saturation.
It is assumed that the desired $x$ position, $x_d$, is chosen sufficiently far along the $x$-axis to give enough time for convergence of the $y$ position and orientation of the robot. 
\begin{figure}[!t]
\centering
\includegraphics[scale=1.5]{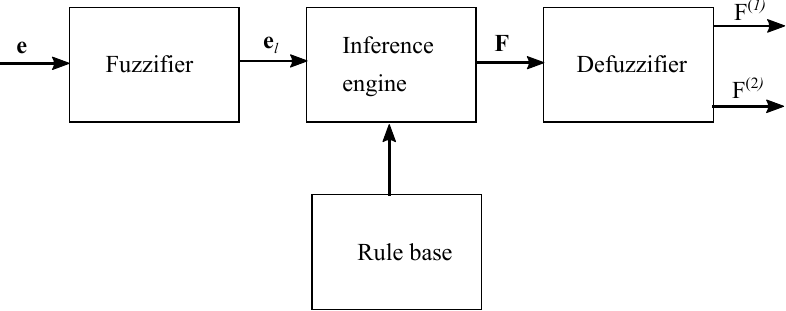}
\caption{The Fuzzy Inference Systems (FIS).}
\label{FuzzyInferenceModefied_Jour}
\end{figure}

The inputs to the FLC are the error in orientation, the error in position along the $y$-axis, and the derivative of the error in orientation.
These inputs are given by the vector $\textbf{e}= [e_{\theta},\:e_{y},\:\dot{e}_{\theta}]^T$ and defined as
\begin{equation}
\begin{aligned}
&e_\theta = \theta_d - \theta,  \\
&e_{y} = {y}_d - y , \\
&\dot{e}_\theta = -\dot{\theta}.
\end{aligned}
\label{eq_13}
\end{equation}
Additionally, the input to the proportional controller is the error in the position along the $x$-axis, given as 
\begin{equation}
\begin{aligned}
&e_x = x_d - x.
\end{aligned}
\end{equation}
 
The Fuzzy Inference System (FIS), a process of mapping from a given input to an output using fuzzy logic, consists of 4 building blocks; 
1) fuzzification, 
2) fuzzy rule-base, 
3) fuzzy inference engine, and 
4) defuzzification. 
A block diagram of this process as applied to the current problem is shown in Fig.~\ref{FuzzyInferenceModefied_Jour}. 
Fuzzification converts the crisp input $\textbf{e}$ to fuzzy input $\textbf{e}_l$ using membership functions to determine the degree of membership.
The fuzzy rule-base is applied to the fuzzified inputs in the fuzzy inference engine and fuzzy outputs are generated. Finally, the fuzzy output is made crisp through defuzzification; in the current work, these are the states $F^{(i)}$ for each brake.

\begin{figure}[h]
\centering
\includegraphics[width=0.75\textwidth]{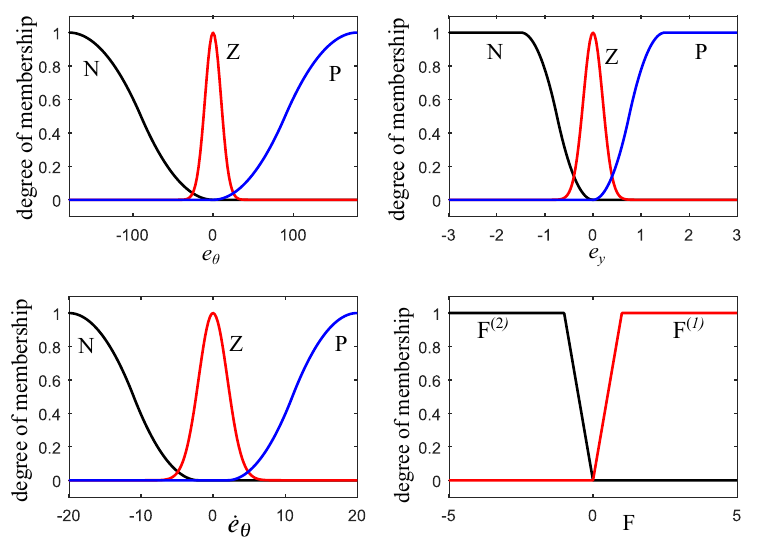}
\caption{Membership functions with their corresponding universe of discourse.}
\label{membership_experiment_Jour}
\end{figure}

The membership functions used in fuzzification of the inputs are \textit{sigma z} for the linguistic variable ``negative'' (N),  \textit{Gaussian} for ``zero'' (Z), and \textit{sigma s} for ``positive'' (P).
The output membership functions used in defuzzification are \textit{sigma z} for activating Brake 2 $( F^{(2)} )$ and  \textit{sigma s} for activating Brake 1 $(F^{(1)})$. 
The membership functions and their universe of discourse for each input variable and the control output are shown in Fig.~\ref{membership_experiment_Jour}. 
Note that the universe of discourse for each membership function together with the percentage of overlap between the membership functions are tunable parameters.
The centroid of area (COA) method is used for defuzzification of the output.
Also note that the defuzzified output is traditionally a crisp, continuous variable but here the values of $F^{(1)}$ and $F^{(2)}$ are discrete, representing the discrete states of the brakes. Algorithm~\ref{alg:the_alg} illustrates the conversion from continuous to discrete braking states for the MAMR.
%
\begin{algorithm}[h]
\begin{algorithmic}
  \caption{Conversion from continuous to discrete braking states}
  \label{alg:the_alg}
  \IF{$F$ is P}
    \STATE $F^{(1)}$ is 1
    \STATE $F^{(2)}$ is 0
  \ELSIF {$F$ is N}
    \STATE $F^{(1)}$ is 0
    \STATE $F^{(2)}$ is 1 
  \ELSE  
    \STATE $F^{(1)}$ is 0
  \STATE $F^{(2)}$ is 0 
  \ENDIF
\end{algorithmic}
\end{algorithm}


The fuzzy rule-base used for generating the fuzzy output from the fuzzy input is illustrated in Table~\ref{FuzzyRulebase}. 
The rule-base contains a number of $IF-THEN$ rules that relate the consequent control action to the antecedent inputs. 
The dashed entry in Table~\ref{FuzzyRulebase}, is the case where both the $y$ position and orientation converge to the desired values, resulting in both brakes being off, as handled in Algorithm~\ref{alg:the_alg}.
\begin{table}[h]
\centering
\caption{\selectfont The fuzzy rule-base for parking the MAMR}
\label{FuzzyRulebase}
\begin{tabular}{|c|c|c|c||c|c|c||c|c|c| }
\hline
$e_{y}$ & \multicolumn{3}{c|}{N} & \multicolumn{3}{c|}{Z} & \multicolumn{3}{c|}{P} \\ \hline
\backslashbox{$e_\theta$}{$\dot{e}_\theta$}     & N      & Z     & P     & N      & Z     & P     & N      & Z     & P  \\ \hline
N    &P        &N       &N       &P        &N      &N       &P       &N       &N  \\ \hline
Z    &N        &N       &P       &N        &-      &P       &N       &P       &P  \\ \hline
P    &N        &P       &P       &N        &P      &P       &N       &P       &P  \\ \hline
\end{tabular}
\end{table}



\section{Experimental Setup}
\label{experimental_setup}
In the following subsections we provide an overview of our experimental setup to include the MAMR prototype and hardware overview.
\subsection{Prototype Design} \label{prototype}
\begin{figure}[!t]
\centering
\includegraphics[scale=1.4]{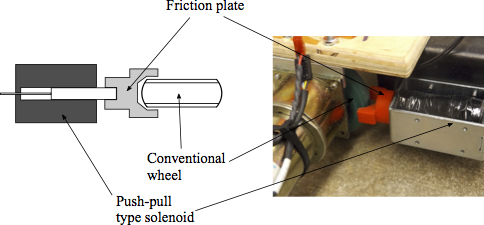}
\caption{Braking actuator with conventional wheel.}
\label{BrakingActuator}
\end{figure}

There are many ways one could conceive of a braking actuator and, likewise, there are innumerable ways in which a MAMR could be configured.  In our previous work~\cite{simmons2016mechatronic}, we presented the design of a ball-type caster that was used for the braking actuator for the MAMR.
One of the limitations of this design however was the limitation of operation being constrained to indoor use on a smooth surface.  In the current work, the ball-caster actuation is replaced with free rolling wheels (un-actuated) with an actual braking mechanism similar to an automobile or bicycle as illustrated in Fig.~\ref{MAMR_II_Jour}.  This new design is advantageous in that an improvement in frictional force between the wheel and ground is obtained while decreasing the overall mechanical complexity.  
A mechanical drawing and prototype of the design (referred to as brake with conventional wheel from now on) is illustrated in Fig.~\ref{BrakingActuator}. 

To brake the conventional wheel, a push-type solenoid acting along the wheel plane is used.
When activated, the solenoid pushes the friction plate onto the outer radius of the wheel which prevents the wheel from rotating.  This is similar to disc brakes used in cars, except the braking force is parallel to the wheel plane as opposed to the perpendicular braking force in disc brakes.
Note that if the brakes are not activated, the conventional wheels spin freely in the direction of motion.
We note that using conventional wheels as opposed to ball-caster type wheels does constrain the robot motion to the $y_r$ direction, i.e., no lateral motion can occur.  
As a result, the robot in the current work is nonholonomic, having less controllable degrees of freedom than the total degrees of freedom of the system.
\begin{figure}[!t]
\centering
\includegraphics[scale=1.5]{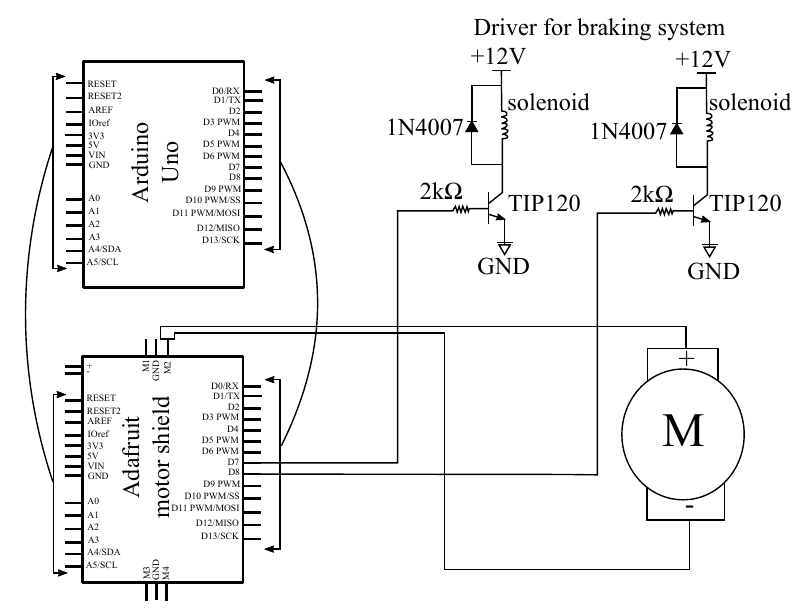}
\caption{The schematics for the electronics of the MAMR.}
\label{Electronics}
\end{figure}
%
\subsection{Hardware Overview} \label{hardware_overview}
The schematic for the electronics used in the MAMR is shown in Fig.~\ref{Electronics}, which includes an Arduino UNO microcontroller, an Adafruit motor driver to drive the omni-directional drive wheel, and a braking actuator driver to activate each solenoid.
The braking actuator driver is an NPN transistor circuit which uses TIP120 transistors with a base resistance of 2k$\Omega$ and a fly-back diode to prevent voltage spikes due to PWM switching across the brake inductance.
The omni-directional drive wheel is powered by a SHR3.6 12V battery and the braking actuator by a 12V Lipo battery pack. 
The rest of the electronics are powered by 8 AA batteries.

The experimental setup for implementing the controller on the hardware includes the MAMR prototype shown in Fig.~\ref{MAMR_II_Jour},
a motion capture system (used for state estimation in the current work), and a desktop computer with \textsc{Matlab}\textsuperscript{\textregistered}~\footnote{MATLAB and SIMULINK are registered trademarks of The Mathworks, Inc., Natick, MA,
USA.} installed.

\section{Experimental Results} \label{experimental_result}
%
\begin{figure}[!t]
\centering
\includegraphics[scale=1.5]{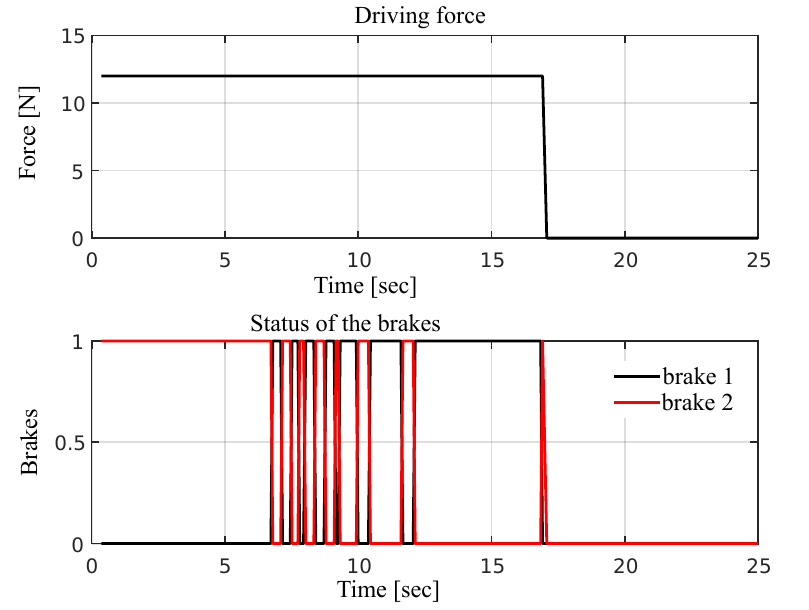}
\caption{ The result for the driving force (top) and status of the brakes (bottom). The driving force and the brakes are set to zero once the $y$ position and orientation converge to the desired values.}
\label{park_exper_simu_control}
\end{figure}
In this section, the experimental results for the parking problem defined in Fig.~\ref{ParkingProblem_MAMR2} are presented. 
The results are presented in the following two subsections illustrating two different scenarios, namely:
\begin{enumerate}
	\item fuzzy logic control in which we drive the MAMR to a desired final configuration $\mathcal{C}_d = [-, y_d,\theta_d]$ where $x_d$ is free and 
	\item parking control in which we drive the MAMR to a full final configuration $\mathcal{C}_d = [x_d,y_d,\theta_d]$.
\end{enumerate}
The percentage of overlap of the membership functions and the admissible
ranges are used as the design parameters to tune the rate of convergence to $\mathcal{C}_d$ in either scenario.  Such parameter tuning is required not for performance issues, but rather to account for limitations in the volume size of the motion capture system used for state estimation. 
\begin{figure}[!t]
\centering
\includegraphics[scale=1.5]{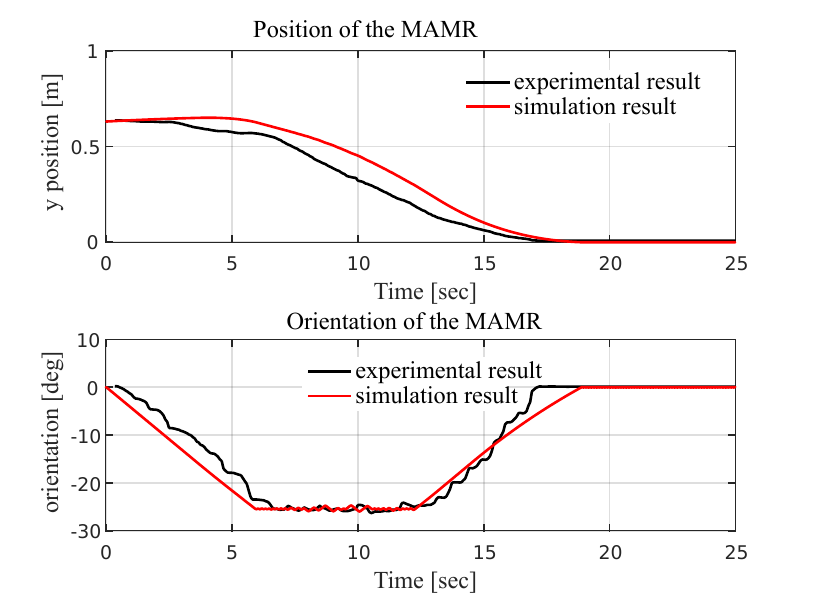}
\caption{FLC results: $y$ position of the robot (top), and orientation of the robot (bottom).}
\label{park_exper_simu_configuration}
\end{figure}
\subsection{Fuzzy Logic Control} \label{FLC_control}
The results for driving the MAMR from an initial configuration $\mathcal{C}_0 = [x_0, y_0, \theta_0]$ to a desired final configuration $\mathcal{C}_d = [-, y_d, \theta_d]$ with $x_d$ free are presented in Figs.~\ref{park_exper_simu_control} and~\ref{park_exper_simu_configuration}.  In addition, for completeness, both simulation results and experimental data are presented in the figures.  In this case, once the MAMR converges to the desired final configuration $\mathcal{C}_d$, i.e., desired $y$ position and the orientation, the MAMR is stopped manually.

The parameters used for simulation are given in Table~\ref{Parameters}. 
The value for the kinetic coefficient of friction was used for a rubber wheel on a dry asphalt floor. In this case, the driving force was kept constant.
The robot's initial configuration is defined as $\mathcal{C}_0 = [0.00,0.63,0.00]$, where again we remind the reader that the units on $\mathbb{R}^2$ are in meters and the units on $\mathbb{S}^1$ are in degrees.  The desired final configuration is $\mathcal{C}_d = [-,0.00,0.00]$ with $x_d$ free, i.e., the $x$ position of the robot is not controlled in this case.
 
As the robot's initial $y$ position is positive, the brake closer to the $x$-axis, i.e.,  Brake 2, is activated and the robot orients about Brake 2 at the initial point.  
The robot takes about 20 seconds to park along the $x$-axis.  
As the final $x$ position is not controlled, the driving force and the brakes are set to zero once the robot is at its desired configuration $\mathcal{C}_d$ as shown in Fig.~\ref{park_exper_simu_control}, causing the robot to stop.  
From the figures, it can be seen that the experimental results align well with the simulation results. The small mismatch in results between the physical platform and the simulation shown in Fig.~\ref{park_exper_simu_configuration} could be due to inaccurate values for the inertia and other parameters assumed in this case. A parameter estimation on these parameters could result in a more accurate result, however this was not carried out in the current work. 
%
\begin{table}[h]
\centering
\caption{The parameters used for simulation results}
\label{Parameters}
\begin{tabular}{llrr}
\hline
Variable    & Description & Value & Units \\
\hline
$m$         & Robot mass                       & 6.8      	& $kg$    \\ 
$I$       & Robot mass moment of inertia       & 1       		& $kgm^2$ \\ 
$\mu^{(i)}_{k}$ & Coefficient of friction at point $i$           & 0.46     	&      \\ 
$x^{(i)}_r$ &\begin{tabular}[x]{@{}l@{}}Distance between the COM to \\braking point $i =1,2$ along the $x_r$ direction \end{tabular}            &0.93  &$m$  \\ 
$y^{(1)}_r$ &\begin{tabular}[x]{@{}l@{}}Distance between the COM to \\braking point along the $y_r$ direction, $i=1$\end{tabular}      &0.155  &$m$    \\ 
$y^{(2)}_r$ &\begin{tabular}[x]{@{}l@{}}Distance between the COM to \\braking point along the $y_r$ direction, $i=2$\end{tabular}      &-0.155  &$m$  \\ 
\hline
\end{tabular}%
\end{table}
%

\begin{figure}[!t]
\centering
\includegraphics[scale=1.7]{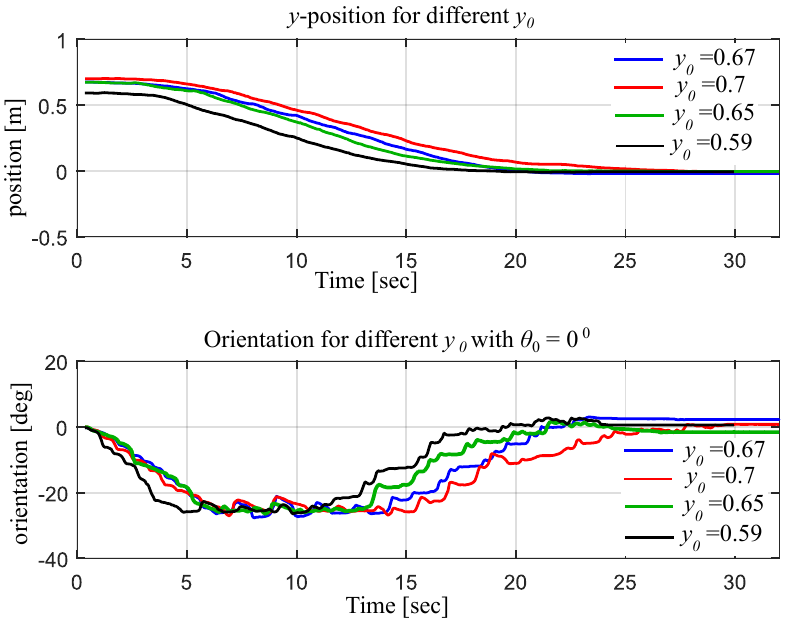}
\caption{Experimental results for different initial conditions; $y$ positions (top) and orientations (bottom).}
\label{position_multi_IC}
\end{figure}

To evaluate the convergence of the MAMR for several different initial configurations $\mathcal{C}_0$, different experiments were performed.  
The resulting $y$ position and orientation for each of the different initial configurations are illustrated in Fig.~\ref{position_multi_IC}.  Because the initial position and orientation are limited by the size of the capture volume, the experiments were performed by changing the initial $y$ position of the MAMR, $y_0$, while the initial orientation of the MAMR, $\theta_0$, was set to zero degrees.  While the experiments were limited by the motion capture volume, the authors have no reason to believe that the robot would not perform similarly over a larger operating region. 
%
\subsection{Parking Control} \label{parking_control}
\begin{figure}[!t]
\centering
\includegraphics[scale=1.7]{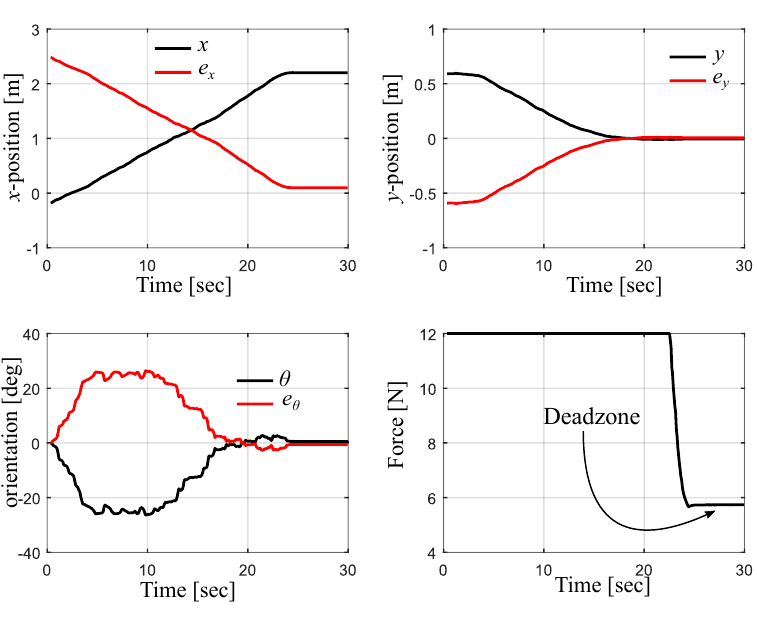}
\caption{Experimental results for parking control.}
\label{Full_parking_result}
\end{figure}

In this section experimental results for parking control in which the MAMR is driven to the desired final configuration of $\mathcal{C}_d = [x_d, y_d, \theta_d]$ from a given initial configuration of  $\mathcal{C}_0 = [x_0, y_0, \theta_0]$ are presented. Fig \ref{Full_parking_result} shows a typical result.  
The robot's initial configuration is defined as $\mathcal{C}_0=[-0.20,0.60,0.00]$ and the desired final configuration $\mathcal{C}_d=[2.30,0.00,0.00]$.
From Fig.~\ref{Full_parking_result}, it can be seen that the driving force remains constant during the convergence of the $y$ position and orientation, and then becomes a function of the error in $x$ position.
The $y$ position and orientation converge in about $18$ seconds and then the robot starts to correct for the error in $x$ position and it takes another $7$ seconds to park the MAMR to the desired final configuration $\mathcal{C}_d$.
The driving force was kept at some minimum value to prevent the motor from entering the deadzone, i.e., if the driving force remains a function of the error in $x$ position, the driving force will drop below the minimum force required to overcome the static friction in the motor. 
This resulted an offset in the driving force as shown in the bottom right plot of Fig.~\ref{Full_parking_result}.

\begin{figure}[!t]
\centering
\includegraphics[scale=1.7]{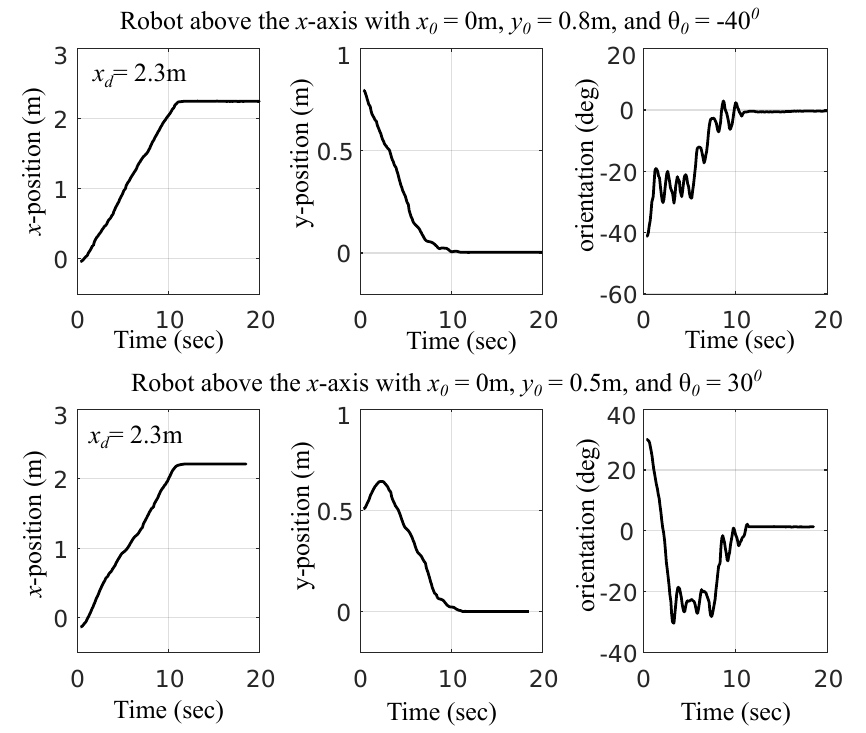}
\caption{Experimental results for the robot with positive initial $y$ position.}
\label{Exper_FLC_positive_Y_pos}
\end{figure}

\begin{figure}[!t]
\centering
\includegraphics[scale=1.7]{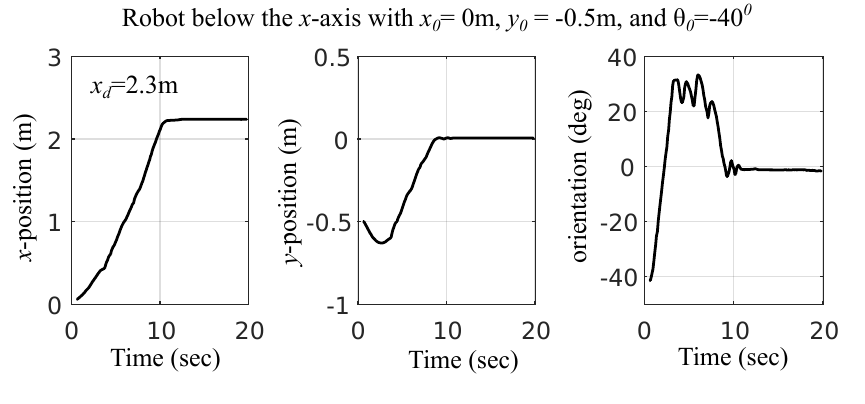}
\caption{Experimental results for the robot with negative initial $y$ position.}
\label{Exper_FLC_negative_Y_pos}
\end{figure}

Based on the dynamics in Section~\ref{modeling}, it is clear that the magnitude of the friction force affects the control authority, and hence the convergence rate.
To improve the convergence rate, the hardware was modified to increase the coefficient of friction between the wheels and ground.
The overall weight of the robot was increased and the ground was painted with skid-resistant paint to get more traction. 
The results with positive initial $y$ position are shown in Fig.~\ref{Exper_FLC_positive_Y_pos}. 
The top three plots are for an initial configuration of $\mathcal{C}_0=[0.00,0.80,-40.00]$, and the bottom three plots are for an initial configuration of $\mathcal{C}_0=[0.00,0.50,30.00]$.
In both cases, the desired final configuration was given by $\mathcal{C}_d=[2.30,0.00,0.00]$.
It can be seen that the robot converges to the desired configuration faster than the results given in Fig.~\ref{Full_parking_result}. 
Fig.~\ref{Exper_FLC_negative_Y_pos} shows results for negative initial $y$ position with an initial configuration defined by $\mathcal{C}_0=[0.00,-0.50,-40.00]$.
As can be seen, the robot converges for a negative initial $y$ position as well.
In all cases, the convergence is improved by improving the coefficient of friction.

\begin{figure}[!t]
\centering
\includegraphics[scale=1.7]{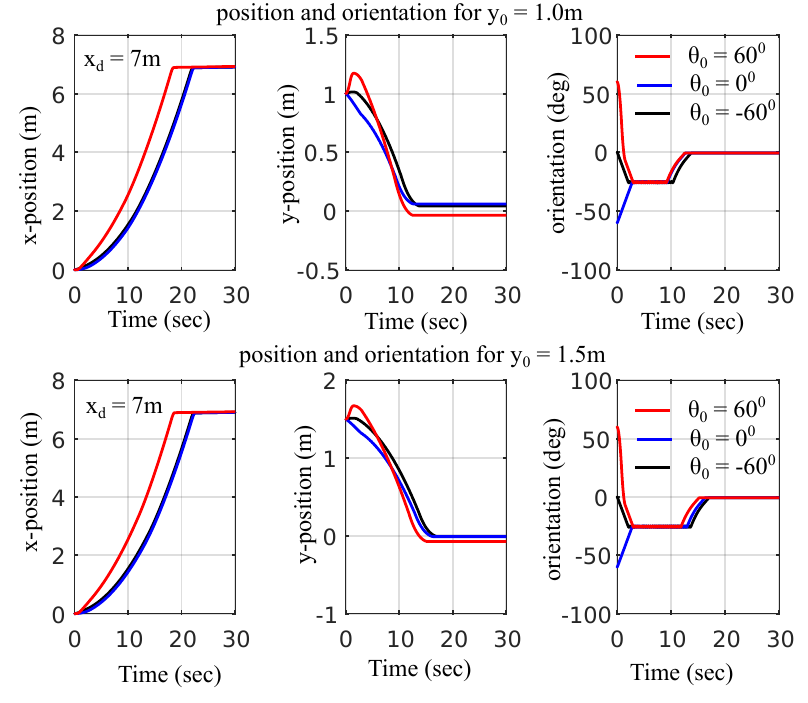}
\caption{Simulation result for different initial conditions.}
\label{FullParking_Simulation}
\end{figure}

Fig.~\ref{FullParking_Simulation} shows the simulation results that test convergence for different initial conditions within a larger operating region. 
The results are for initial configurations of $\mathcal{C}_0=[0.00,1.00,\theta_0]$ and $\mathcal{C}_0=[0.00,1.50,\theta_0]$, where $\theta_0 \in \left\{ 0^\circ, -60^\circ, 60^\circ\right\}$.
In all cases, the robot is made to park at a final configuration of $\mathcal{C}_d=[7.00,0.00,0.00]$ . 
Note that the same convergence can be attained for negative initial $y$ positions as the rule bases are symmetric.

\begin{figure}[!t]
\centering
\includegraphics[width=\columnwidth]{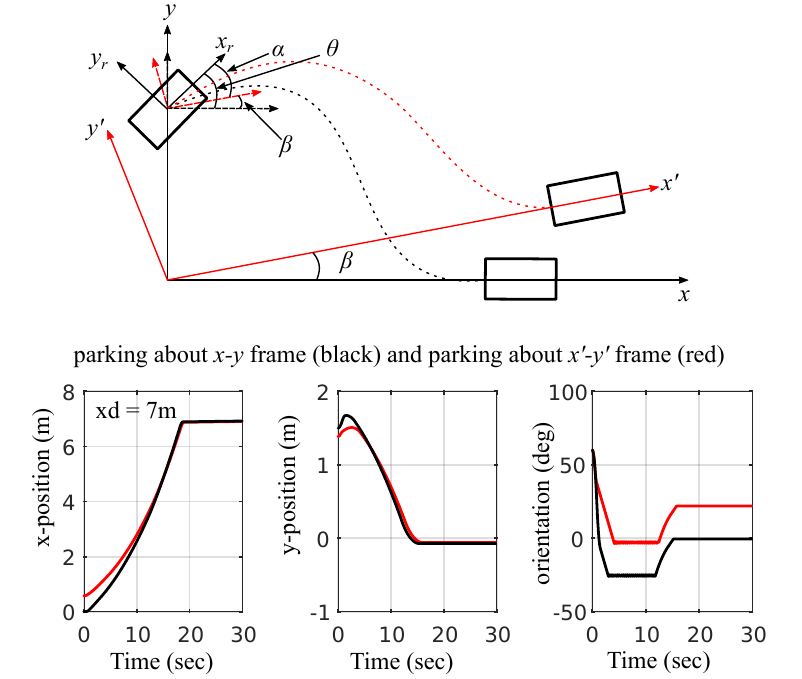}
\caption{Simulation result for parking the robot about $x'$-axis.}
\label{FullParking_Simulation_Any_Tdesired}
\end{figure}

All results so far have demonstrated parking to an angle that aligns with the global $x$ axis.
If the desired final angle is chosen as $\beta^\circ$ from the $x$-axis, parking of the robot can be attained by simply choosing an appropriate global coordinate frame ($x', y'$). 
The angle to correct for the new global coordinate is given by $\alpha = \theta - \beta$. 
Fig.~\ref{FullParking_Simulation_Any_Tdesired} shows the problem description (top) and simulation results (bottom three) for the convergence of the robot about the $x$-axis and $x'$-axis. 
The simulation results are for $\beta = 0^\circ$ (black line), and $\beta = 22^\circ$ (red line) for the same initial configuration. 
%

%


\section{Conclusions and Future Work} \label{conclusion_recommendation}
The current paper presented a new approach for actuating mobile robots during locomotion, that uses a combination of conventional and braking actuators.  The use of the conventional wheels together with brakes results in a nonholonomic robot where the robot is constrained from moving sideways.  The experimental results illustrate that FLC was able to park the MAMR from a given initial configuration to desired final configuration in finite time.  

In the current setup, the motion volume is limited by the capture volume of the motion capture system.  In future work, we would like to investigate a larger space for the motion capture volume to validate the controller under larger initial and final configurations.  We would also like to evaluate whether the experimental system can be better approximated by considering a more complex friction model with stick-slip effects as opposed to Coulomb friction.  Finally, we would like to work on regaining the controllability of the system under actuator failure, which is one of the main motivations for studying the mixed conventional/braking actuation system.

 The authors would also experimentally validate the Sliding Mode Controller (SMC) which was designed in to solve the problem associated with increased fuzzy rule bases if the $x$ and $y$ positions were controlled simultaneously while operating under variable driving force.




\bibliographystyle{IEEEtran}
\bibliography{MAMR_FLC_2018}

\begin{thebibliography}{10}
\providecommand{\url}[1]{#1}
\csname url@samestyle\endcsname
\providecommand{\newblock}{\relax}
\providecommand{\bibinfo}[2]{#2}
\providecommand{\BIBentrySTDinterwordspacing}{\spaceskip=0pt\relax}
\providecommand{\BIBentryALTinterwordstretchfactor}{4}
\providecommand{\BIBentryALTinterwordspacing}{\spaceskip=\fontdimen2\font plus
\BIBentryALTinterwordstretchfactor\fontdimen3\font minus
  \fontdimen4\font\relax}
\providecommand{\BIBforeignlanguage}[2]{{%
\expandafter\ifx\csname l@#1\endcsname\relax
\typeout{** WARNING: IEEEtran.bst: No hyphenation pattern has been}%
\typeout{** loaded for the language `#1'. Using the pattern for}%
\typeout{** the default language instead.}%
\else
\language=\csname l@#1\endcsname
\fi
#2}}
\providecommand{\BIBdecl}{\relax}
\BIBdecl

\bibitem{schneier2015literature}
M.~Schneier and R.~Bostelman, ``Literature review of mobile robots for
  manufacturing,'' \emph{National Institute of Standards and Technology,
  NewYork}, 2015.

\bibitem{siegwart2011introduction}
R.~Siegwart, I.~R. Nourbakhsh, and D.~Scaramuzza, \emph{Introduction to
  autonomous mobile robots}.\hskip 1em plus 0.5em minus 0.4em\relax MIT press,
  2011.

\bibitem{dudek2005visually}
G.~Dudek, M.~Jenkin, C.~Prahacs, A.~Hogue, J.~Sattar, P.~Giguere, A.~German,
  H.~Liu, S.~Saunderson, A.~Ripsman \emph{et~al.}, ``A visually guided swimming
  robot,'' in \emph{IEEE/RSJ International Conference on Intelligent Robots and
  Systems (IROS)}.\hskip 1em plus 0.5em minus 0.4em\relax IEEE, 2005, pp.
  3604--3609.

\bibitem{dudek2007aqua}
G.~Dudek, P.~Giguere, C.~Prahacs, S.~Saunderson, J.~Sattar, L.-A.
  Torres-Mendez, M.~Jenkin, A.~German, A.~Hogue, A.~Ripsman \emph{et~al.},
  ``Aqua: An amphibious autonomous robot,'' \emph{AQUA}, vol.~10, p.~43, 2007.

\bibitem{bouabdallah2007design}
S.~Bouabdallah and R.~Siegwart, ``Design and control of a miniature
  quadrotor,'' in \emph{Advances in unmanned aerial vehicles}.\hskip 1em plus
  0.5em minus 0.4em\relax Springer, 2007.

\bibitem{cutler2012design}
M.~J. Cutler, ``Design and control of an autonomous variable-pitch quadrotor
  helicopter,'' Ph.D. dissertation, Brigham Young University, 2012.

\bibitem{machado2006overview}
J.~T. Machado and M.~F. Silva, ``An overview of legged robots,'' in
  \emph{International Symposium on Mathematical Methods in Engineering}, 2006.

\bibitem{deshmukh2006robot}
A.~Deshmukh and C.~Amarnath, ``Robot leg mechanisms,'' in \emph{B. Tech.
  Seminar Report, Roll}.\hskip 1em plus 0.5em minus 0.4em\relax Citeseer, 2006.

\bibitem{iverach2012ice}
C.~Iverach-Brereton, A.~Winton, and J.~Baltes, ``Ice skating humanoid robot,''
  in \emph{Advances in Autonomous Robotics}.\hskip 1em plus 0.5em minus
  0.4em\relax Springer, 2012.

\bibitem{stepan2009acroboter}
G.~Stepan, A.~Toth, L.~Kovacs, G.~Bolmsjo, G.~Nikoleris, D.~Surdilovic,
  A.~Conrad, A.~Gasteratos, N.~Kyriakoulis, D.~Chrysostomou \emph{et~al.},
  ``Acroboter: a ceiling based crawling, hoisting and swinging service robot
  platform,'' in \emph{Beyond gray droids: domestic robot design for the 21st
  century workshop at HCI}, vol. 2009, 2009, p.~2.

\bibitem{wang2008biological}
M.~Wang, X.-z. Zang, J.-z. Fan, and J.~Zhao, ``Biological jumping mechanism
  analysis and modeling for frog robot,'' \emph{Journal of Bionic Engineering},
  vol.~5, no.~3, pp. 181--188, 2008.

\bibitem{menon2004gecko}
C.~Menon, M.~Murphy, and M.~Sitti, ``Gecko inspired surface climbing robots,''
  in \emph{IEEE International Conference on Robotics and Biomimetics (ROBIO)},
  2004, pp. 431--436.

\bibitem{art:Jun2018}
J.~Fu, F.~Tian, T.~Chai, Y.~Jing, Z.~Li, and C.~Y. Su, ``Motion tracking
  control design for a class of nonholonomic mobile robot systems,'' \emph{IEEE
  Transactions on Systems, Man, and Cybernetics: Systems}, pp. 1--7, 2018.

\bibitem{art:Li2017}
M.~Li, Y.~Li, S.~S. Ge, and T.~H. Lee, ``Adaptive control of robotic
  manipulators with unified motion constraints,'' \emph{IEEE Transactions on
  Systems, Man, and Cybernetics: Systems}, vol.~47, no.~1, pp. 184--194, Jan
  2017.

\bibitem{art:Chen2017}
C.~Chen, Z.~Liu, Y.~Zhang, and S.~Xie, ``Coordinated motion/force control of
  multiarm robot with unknown sensor nonlinearity and manipulated object's
  uncertainty,'' \emph{IEEE Transactions on Systems, Man, and Cybernetics:
  Systems}, vol.~47, no.~7, pp. 1123--1134, July 2017.

\bibitem{andani2018fuzzy}
M.~T. Andani, S.~Shahmiri, H.~Pourgharibshahi, K.~Yousefpour, and M.~H. Imani,
  ``Fuzzy-based sliding mode control and sliding mode control of a spherical
  robot,'' in \emph{IECON 2018-44th Annual Conference of the IEEE Industrial
  Electronics Society}.\hskip 1em plus 0.5em minus 0.4em\relax IEEE, 2018, pp.
  2534--2539.

\bibitem{morin2008motion}
P.~Morin and C.~Samson, ``Motion control of wheeled mobile robots,'' in
  \emph{Springer Handbook of Robotics}.\hskip 1em plus 0.5em minus 0.4em\relax
  Springer, 2008, pp. 799--826.

\bibitem{de2001control}
A.~De~Luca, G.~Oriolo, and M.~Vendittelli, ``Control of wheeled mobile robots:
  An experimental overview,'' in \emph{Ramsete}.\hskip 1em plus 0.5em minus
  0.4em\relax Springer, 2001.

\bibitem{tzafestas2013introduction}
S.~G. Tzafestas, \emph{Introduction to mobile robot control}.\hskip 1em plus
  0.5em minus 0.4em\relax Elsevier, 2013.

\bibitem{mavroidis2005}
C.~Mavroidis, C.~Pfeiffer, and M.~Mosley, ``conventional actuators, shape
  memory alloys, and electrorheological fluids.''

\bibitem{pratt1995series}
G.~A. Pratt and M.~M. Williamson, ``Series elastic actuators,'' in \emph{IEEE
  International Conference on Intelligent Robots and Systems (IROS)}, vol.~1,
  1995.

\bibitem{pratt2004series}
J.~E. Pratt and B.~T. Krupp, ``Series elastic actuators for legged robots,'' in
  \emph{Unmanned Ground Vehicle Technology}, vol. 5422, 2004, pp. 135--145.

\bibitem{litman2014autonomous}
T.~Litman, ``Autonomous vehicle implementation predictions,'' \emph{Victoria
  Transport Policy Institute}, vol.~28, 2014.

\bibitem{bedillion2014distributed}
M.~Bedillion, R.~Hoover, and J.~McGough, ``A distributed manipulation concept
  using selective braking,'' in \emph{American Control Conference (ACC)}, 2014,
  pp. 3322--3328.

\bibitem{framstad2015control}
E.~Framstad and M.~D. Bedillion, ``Control for robots with braking actuators in
  a uniform force field,'' in \emph{ASME 2015 International Mechanical
  Engineering Congress and Exposition}, 2015.

\bibitem{nikshi2018trajectory}
W.~M. Nikshi, R.~C. Hoover, M.~D. Bedillion, and S.~Shahmiri, ``Trajectory
  tracking of the mixed conventional/braking actuation mobile robots using
  model predictive control,'' in \emph{2018 IEEE 14th International Conference
  on Control and Automation (ICCA)}.\hskip 1em plus 0.5em minus 0.4em\relax
  IEEE, 2018, pp. 704--709.

\bibitem{art:zadah73}
L.~A. Zadeh, ``Outline of a new approach to the analysis of complex systems and
  decision processes,'' \emph{IEEE Transactions on Systems, Man, and
  Cybernetics}, vol. SMC-3, pp. 28--44, 1973.

\bibitem{castillo2006fuzzy}
O.~Castillo, L.~T. Aguilar, and S.~C{\'a}rdenas, ``Fuzzy logic tracking control
  for unicycle mobile robots.'' \emph{Engineering Letters}, vol.~13, no.~2, pp.
  73--77, 2006.

\bibitem{hoover2005hybrid}
R.~C. Hoover, M.~P. Schoen, and D.~S. Naidu, ``Hybrid computing techniques for
  collaborative control of ucavs,'' in \emph{ASME 2005 International Mechanical
  Engineering Congress and Exposition}, 2005, pp. 151--160.

\bibitem{hoover2005fusion}
R.~Hoover, M.~Schoen, and S.~Naidu, ``Fusion of hard and soft control for
  uninhabited aerial vehicles,'' in \emph{IFAC 16th Annual World Congress},
  2005.

\bibitem{rashid2010fuzzy}
R.~Rashid, I.~Elamvazuthi, M.~Begam, and M.~Arrofiq, ``Fuzzy-based navigation
  and control of a non-holonomic mobile robot,'' \emph{arXiv preprint
  arXiv:1003.4081}, 2010.

\bibitem{simmons2016mechatronic}
J.~W. Simmons, W.~M. Nikshi, M.~D. Bedillion, and R.~C. Hoover, ``Mechatronic
  design of a mixed conventional/braking actuation mobile robot,'' in
  \emph{ASME 2016 International Mechanical Engineering Congress and
  Exposition}, 2016.

\bibitem{nikshi2016parking}
W.~M. Nikshi, M.~D. Bedillion, and R.~C. Hoover, ``Parking control of mixed
  conventional/braking actuation mobile robots using fuzzy logic control,'' in
  \emph{ASME, International Mechanical Engineering Congress and Exposition},
  2016.

\bibitem{watanabe1998control}
K.~Watanabe, ``Control of an omnidirectional mobile robot,'' in
  \emph{Knowledge-Based Intelligent Electronic Systems, 1998. Proceedings
  KES'98. 1998 Second International Conference}, vol.~1, 1998.

\end{thebibliography}

\end{document}